# Analysis and Synthesis Prior Greedy Algorithms for Non-linear Sparse Recovery

Kavya Gupta[*], Ankita Raj[+] and Angshul Majumdar[#]

| [*]*IIIT Delhi* | [+]*IIIT Delhi* | [#]*IIIT Delhi* |
| *Okhla Phase 3* | *Okhla Phase 3* | *Okhla Phase 3* |
| *Delhi, 110020, India* | *Delhi, 110020, India* | *Delhi, 110020, India* |
| *kavya1482@iiitd.ac.in* | *ankita1477@iiitd.ac.in* | *angshul@iiitd.ac.in* |

*Abstract*: In this work we address the problem of recovering sparse solutions to non-linear inverse problems. We look at two variants of the basic problem – the synthesis prior problem when the solution is sparse and the analysis prior problem where the solution is co-sparse in some linear basis. For the first problem, we propose non-linear variants of the Orthogonal Matching Pursuit (OMP) and CoSamp algorithms; for the second problem we propose a non-linear variant of the Greedy Analysis Pursuit (GAP) algorithm. We empirically test the success rates of our algorithms on exponential and logarithmic functions. We model speckle denoising as a non-linear sparse recovery problem and apply our technique to solve it. Results show that our method outperforms state-of-the-art methods in ultrasound speckle denoising.

## 1. Introduction

In the last decade, there has been a lot of interest in solving linear inverse problems where the solution is known to be sparse, especially when the problem is under-determined. In his seminal work Donoho [1] showed that sparse solutions are necessarily unique, even when it is under-determined.

There are two approaches to solve the sparse recovery problem. One approach is to relax the NP hard $l_0$-norm by its nearest convex surrogate the $l_1$-norm. Such $l_1$-norm minimization problems can be solved using linear programming. The other approach to solve the sparse recovery problem is via greedy methods. These techniques are based on iteratively finding the support (via some heuristic) of the solution and estimating the values at those positions. Orthogonal Matching Pursuit (OMP) [2] and CoSamp [3] are the two popular greedy recovery algorithms.

There is a variant to the basic problem (2), where the solution is not sparse but has a co-sparse representation in a transform domain. This is the analysis prior problem [4]. Much like the synthesis prior problem, there are two ways to solve the NP hard analysis prior problem – via convex relaxation and via greedy algorithms like Greedy Analysis Prior (GAP).

In this work, our interest is in solving non-linear sparse recovery problems. In a previous study [5], a non-linear $l_1$-norm minimization problem was proposed to solve it. The problem with such an approach is its slow convergence (owing to small step-size). To overcome the limitations of speed, we propose greedy algorithms based on the modifications of OMP and CoSamp to solve the synthesis prior non-linear sparse recovery problem (when the solution is sparse) and modification of GAP to solve the analysis prior non-linear co-sparse recovery problem (when the solution is co-sparse in a transform domain). Before going into our proposed techniques, we will discuss a practical motivation of such non-linear sparse recovery problems in the next section; we model ultrasound speckle denoising as a non-linear sparse recovery problem.

## 2. Speckle Denoising in Ultrasound

Ultrasound images are corrupted by speckle noise [6-10]. Speckle noise is multiplicative in nature and can be modeled as,

$$x = x_0 \cdot n \tag{1}$$

Here $x_0$ is the clean image (to be recovered), $n$ is the noise and $x$ is the speckle corrupted image.

Broadly there are two classes of approach. The more popular one is to log-transform (1) to an additive noise model and apply wavelet (or other *let) based thresholding method to remove noise, followed by exponential transform to get back the cleaned image [6-9]. The other approach is to apply spatial domain filtering methods like non local means to get rid of the noise [10].

In this work we propose a novel formulation. In a fashion similar to thresholding technique, we first convert (1) to an additive model.

$$\log(x) = \log(x_0) + \log(n) \tag{2}$$

Prior studies like [6-9] and others assumed that $\log(x_0)$ is sparse in a transform domain. They applied the sparsifying transform to $\log(x_0)$ and thresholded the coefficients (strictly speaking the log transform of the noise follows a Fisher-Tipett distribution; but for the purpose of simplicity it has been considered to be white Gaussian in nature). It is proven that wavelet (and other *let) transform is able to sparsely represent piecewise smooth signals like images; however to the best of our knowledge there is no formal proof that says that logarithm of a piecewise signal will also be piecewise smooth and hence sparsely representable in such sparsifying transform domains.

In this work, we follow the standard assumption that that the image is piecewise smooth and hence sparsely representable in a transform domain - this has been used profusely in removing additive Gaussian noise from images and in Compressed Sensing based additive noise removal. In the most general case we assume that $H$ is the sparsifying transform (not necessarily orthogonal or tight-frame). This allows formulating speckle denoising as follows:

$$\min_{x_0} \|\log(x) - \log(x_0)\|_2^2 + \lambda \|Hx_0\|_1 \tag{3}$$

The $l_2$-norm for data fidelity has been widely used in speckle denoising; studies like [6-10] are based on the assumption that $\log(n)$ follows a Gaussian distribution - hence the justification of using the Euclidean norm.

This is a typical candidate for analysis prior non-linear co-sparse recovery. If we assume that the sparsifying transform is orthogonal (or tight framed) we can pose denoising as a sparse synthesis prior problem:

$$\min_{\alpha} \left\| \log(x) - \log(H^T \alpha) \right\|_2^2 + \lambda \left\| \alpha \right\|_1 \tag{4}$$

There are no off-the shelf algorithms for solving the sparse recovery problem from non-linear measurements. In the next section, we will discuss the few studies on this topic.

## 3. Literature Review

Literature on non-linear sparse recovery is parsimonious. There are two studies [11, 12] of theoretical nature that explores the conditions under which such recovery is possible. They do not propose practical algorithms for solving such problems. To the best of our knowledge, there is only a single work [13] from 2008 that proposed a greedy algorithm (variant of OMP) to solve the synthesis prior non-linear sparse recovery problem; but in essence their algorithm recovered a sparse solution to a linear problem where the cost function is not Euclidean.

To the best of our knowledge, the only algorithmic work on solving non-linear sparse recovery problems in [5]. There the authors proposed an $l_1$-norm minimization approach; the recovery was posed as (assuming $f(.)$ to be the non-linear function):

$$\min_{x} \left\| y - f(x) \right\|_2^2 + \lambda \left\| x \right\|_1 \tag{5}$$

Solving (5) for very small values of ε is as good as solving the equality constrained problem. The algorithm in [8] is based on the modified Iterative Soft Thresholding Algorithm (ISTA). It is a two step algorithm. The first step is a gradient descent that partially solves the non-linear least squares problem –

$$b = x_{k-1} - \sigma \nabla_x \left\| y - f(x) \right\|_2^2 \Big|_{x = x_{k-1}}$$

where σ is the step-size.

The second step is to threshold b in order to project it on the $l_1$-ball, to yield a sparse solution –

$$x_k = SoftTh(b, \tau)$$

The problem with such an optimization based approach is that it is parametric. For the linear case, well known techniques exist to find the step-size (σ) and the threshold τ; but for non-linear cases finding the values is difficult. In [5], the step-size was determined by the inverse of the Lipchitz bound of the cost function. But it is a very pessimistic (small) step-size and the algorithm converges slowly; also the threshold parameter was found manually. To overcome these limitations we propose greedy algorithms to solve the non-linear sparse recovery problem.

## 4. Proposed Greedy Algorithms

### 4.1. Solving the Sparse Synthesis Prior Problem

Before discussing the non-linear extensions, let us discuss the greedy algorithms for linear sparse recovery. The most commonly used algorithm for the purpose is Orthogonal Matching Pursuit (OMP). The algorithm is as follows:

**OMP Algorithm**

Initialize: Index set - $\Omega = \varnothing$; x=0.
Repeat for k iterations
Compute correlation - $c = abs(A^T(y - Ax))$
Select the index - $i = \arg\max_i c_i$
Update support - $\Omega = \Omega \cup i$
Estimate non-zero values - $x_\Omega = \min_x \|y - A_\Omega x\|_2^2$
Finally impute other indices in x with zeroes.

There are two issues with the OMP algorithm. First, it only selects one index of the support set in every iteration. This is slow; there are simple techniques to accelerate OMP like Stagewise OMP or Stagewise Weak OMP. However, slow selection is not the most pressing problem. The bigger problem is that an index once selected, remains; OMP (or its aforesaid variants) cannot prune an incorrectly selected index at a later iteration. The CoSamp overcomes overcome both these issues. It allows for selection of multiple indices in every iteration (top 2k) and it allows for pruning (by keeping top k).

**CoSamp Algorithm**

Initialize: Index set - $\Omega = \varnothing$; x=0.
Repeat until convergence
Compute correlation - $c = abs(A^T(y - Ax))$
Select the top 2K support from c - $\omega$
Update support - $\Omega = \Omega \cup \omega$
Perform least squares estimate - $b_\Omega = \min_x \|y - A_\Omega x\|$
Update x by pruning b and keeping top K values.
Finally impute other indices in x with zeroes.

Let us look at computing the correlation 'c' for both the algorithms. It is computed as $A^T(y - Ax)$; this is the negative gradient of the Euclidean cost function $\|y - Ax\|_2^2$. We want to solve the linear sparse recovery problem, i.e.

$$x: y = Ax \text{ s.t. } \|x\|_0 = k \tag{6}$$

The equality (y=Ax) holds only at convergence, for all other iterations (in OMP and CoSamp) one finds x by minimizing the Euclidean cost function $\|y - Ax\|_2^2$.

Our proposed algorithm for non-linear sparse recovery is based on the same approach. For our problem, the Euclidean cost function is $\|y - f(x)\|_2^2$. We would like to solve the non-linear sparse recovery problem:

$$x: y = f(x) \text{ s.t. } \|x\|_0 = k \tag{7}$$

Following same argument as in the linear case, our 'correlation' for non-linear problems should be defined as:

$$c = -\nabla_x \|y - f(x)\|_2^2 \tag{8}$$

The rest of the OMP and the CoSamp algorithm remains as it is except for the least squares step. Previously, one needed to solve a linear least squares problem. We need to solve a non-linear least squares problem. But solving non-linear least squares is a well studied area and there is no dearth of efficient algorithms; we use the Levenberg Marquadt method.

Non-Linear OMP Algorithm

Initialize: Index set - $\Omega = \emptyset$; x=0.
Repeat for k iterations
Compute correlation - $c = abs(\nabla_x \|y - f(x)\|_2^2)$
Select the index - $i = \arg\max_i c_i$
Update support - $\Omega = \Omega \cup i$
Estimate non-zero values - $x_\Omega = \min_x \|y - f_\Omega(x)\|$
Finally impute other indices in x with zeroes.

Non-linear CoSamp Algorithm

Initialize: Index set - $\Omega = \emptyset$; x=0.
Repeat until convergence
Compute correlation - $c = abs(\nabla_x \|y - f(x)\|_2^2)$
Select the top 2K support from c - $\omega$
Update support - $\Omega = \Omega \cup \omega$
Perform least squares estimate - $b_\Omega = \min_x \|y - f_\Omega(x)\|$
Update x by pruning b and keeping top K values.
Finally impute other indices in x with zeroes.

## 4.2. Solving the Co-sparse Analysis Prior Problem

There are well known $l_1$-minimization based algorithms to solve the sparse synthesis and co-sparse analysis prior problem for the linear case; there are a few algorithms (mentioned before) for solving the sparse synthesis problem for non-linear measurements, but there is no prior technique to solve the non-linear problem. We propose to modify the Greedy Analysis Prior (GAP) algorithm to solve the co-sparse analysis prior problem for non-linear measurements. GAP starts from a dense solution and in every iteration it prunes one non-zero element from it. After pruning, the solution (at the current co-support) is updated by solving a least squares problem. The algorithm for linear measurements is given below:

GAP Algorithm

Initialize: Index set - $\Lambda = \{1, 2, 3, \ldots, n\}$
$x = \arg\min_x \|\Omega_\Lambda x\|_2^2$ subject to $y = Ax$
Repeat for k iterations
Compute - $c = abs(\Omega x)$
Select the index - $i = \arg\max_i c_i$
Update co-support - $\Lambda = \Lambda \setminus i$
Update solution - $x = \arg\min_x \|\Omega_\Lambda x\|_2^2$ subject to $y = Ax$

We are interested in solving the co-sparse analysis prior problem when the measurement is non-linear. We modify the GAP algorithm to achieve this. The first step is to obtain the sparse representation of the solution x, here it is represented as c; it is computed by applying the sparsifying transform to the current solution. As long as the sparsifying transform is linear (as in our case), this step does not change when the measurement is non-linear. The pruning operation also remains as it is. The only change that needs to be introduced is in solving the least squares problem for updating x. We have a non-linear measurement function therefore we need to solve a least squares problem with non-linear equality constraints. There are several well known methods to solve such problems.

Non-linear GAP Algorithm

Initialize: Index set - $\Lambda = \{1, 2, 3, ....., n\}$
$x = \arg\min_x \|\Omega_\Lambda x\|_2^2$ subject to $y = f(x)$
Repeat for k iterations
Compute - $c = abs(\Omega x)$
Select the index - $i = \arg\max_i c_i$
Update cosupport - $\Lambda = \Lambda \setminus i$
Update solution - $x = \arg\min_x \|\Omega_\Lambda x\|_2^2$ subject to $y = f(x)$

We believe in reproducible research. Matlab implementations of both the algorithms are available at the author's Matlab Central account [14].

## 5. Experimental Results

### 5.1. Experiments on Synthetic Data

There is no benchmark to compare our greedy algorithms non-linear recovery. So we tested them against the linear recovery problems. We used the original algorithms (OMP and CoSaMP) for linear recovery – the results for this problem are well known, vis-à-vis we used our proposed non-linear OMP and CoSaMP algorithms to solve the linear recovery problem. We also used the proposed algorithms to solve the exponential and logarithmic inverse problems to check how they compare against the well-known success rates of linear recovery. The results are shown in Fig. 1 for i.i.d Gaussian and Bernoulli measurements; the size of the problem is 40 X 100. The number of non-zero values (k) in x is varied. For each configuration, 1000 trials are generated; a trial is considered successful if the normalized mean squared error ($NMSE = \frac{\|original - reconstructed\|_2}{\|original\|_2}$) is below $10^{-3}$.

We find that for OMP, our non-linear algorithm performs better than the original for linear recovery problems. But for CoSaMP, the algorithm for linear recovery [2] performs much better than our proposed technique on linear problems. The success rates for the non-linear problems show a similar trend, but are worse than that of their linear counterparts.

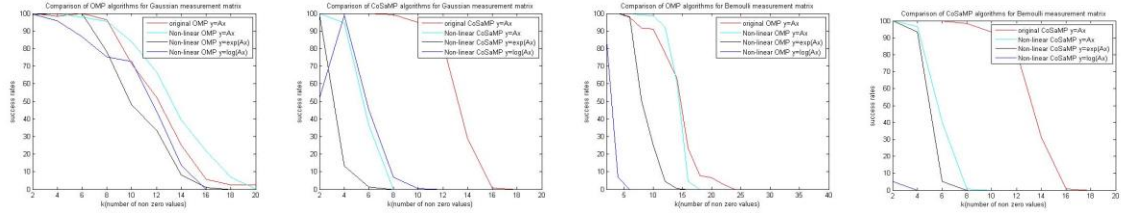

**Figure 1.** Comparison of success rates for various functions with OMP and CoSaMP for i.i.d. Gaussian and Bernoulli measurement matrix.

In the second set of experiments we show how the success rates vary with number of non-zero values in the solution (k) and number of measurements (m) for our proposed non-linear OMP and CoSaMP algorithms for exponential and logarithmic functions. In Fig. 2 we show the results for an i.i.d Gaussian matrix; owing to limitations in space we cannot show the results for Bernoulli measurements, but the trends are similar. For OMP, the results follow a well-known pattern – the success rates decrease with increasing number of non-zeroes in the solution and increases with the number of measurements. The results for CoSaMP on the logarithmic problem shows a peculiar pattern – the success rates increases initially with the increase in the number of non-zero values and then the rate decreases as expected; this pattern is only visible when the number of measurements is 60 and 80. We have tested this scenario multiple times and always saw the same pattern. We do not know how to explain this behavior.

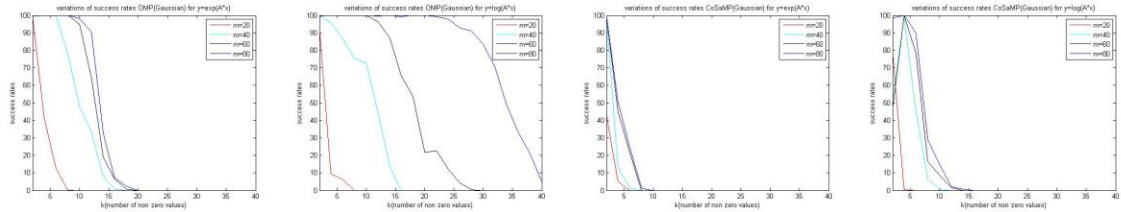

**Figure 2.** Variation of success rates for OMP and CoSaMP with number of non-zero values and number of measurements for i.i.d Gaussian matrix.

So far we have discussed results for the synthesis prior problems. We tested the non-linear GAP algorithm with the standard linear GAP technique for recovering the solution for a linear problem. The results are shown in Fig. 3 (a and b); these pertain to 40% sampling. We find that our algorithm performs slightly worse than the original technique. In the same graph we also compared the results for recovering exponential and logarithmic function with our proposed technique. In Fig. 3 (c and d) we compared our method for recovering exponential and logarithmic functions for non-linear i.i.d Gaussian and measurements. The success rates follow the usual trend.

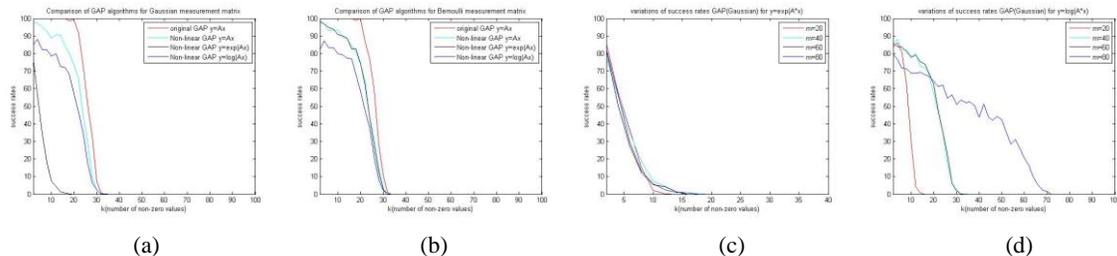

(a)      (b)      (c)      (d)

**Figure 3.** (a) and (b) Comparison of success rates for various functions with GAP for i.i.d Gaussian and Bernoulli measurement matrix. (c) and (d) Variation of success rates for

GAP with number of non-zero values and number of measurements for i.i.d Gaussian matrix.

## 5.2. Experiments on Speckle Denoising

The data used in here has been downloaded from a public access database [15]. The database contains images of common carotid artery (CCA) of ten volunteers (mean age 27.5 ± 3.5 years) with different weight (mean weight 76.5 ± 9.7 kg). Images (usually eight images per volunteer) were acquired with Sonix OP ultrasound scanner with different set-up of depth, gain, time gain compensation (TGC) curve and different linear array transducers. The image database contains 84 B-mode ultrasound images of CCA in longitudinal section. The resolution of images is approximately 390x330px.

We have compared our proposed technique with two state-of-the-art methods [6] (published in 2015) and [16] (published 2014). In these studies, it was shown that they outperform other standard speckle denoising techniques like adaptive filtering [17] and isotropic diffusion [18].

The images are artifically corrupted by speckle noise with various noise levels and denoised using [6], [16] and our three proposed techniques. For simplicity, we assume that the noise is white Gaussian. For our OMP and CoSamp we use Daubechies wavelet (3 level decomposition, 8 vanishing moments) as the sparsifying transform and for the GAP we use total variation prior (finite differencing). We assume that the number of non-zero coefficients is 10% of the total number of coefficients; but iterations may stop if the residual is near about 0.

The average (over all images) PSNR and the SSIM for the different techniques are shown in Table 1. In the table it is understood that our proposed techniques are non-linear version of OMP. CoSamp and GAP, therefore we omit the word 'non linear'.

Table 1. PSNR Values from Various Techniques

| Algorithm | Input PSNR = 14 | | Input PSNR = 12 | | Input PSNR = 10 | |
|---|---|---|---|---|---|---|
| | PSNR | SSIM | PSNR | SSIM | PSNR | SSIM |
| Wavelet + NLM [6] | 33.4 | 0.911 | 28.2 | 0.856 | 23.0 | 0.726 |
| Framelet Diffusion [16] | 32.2 | 0.903 | 27.6 | 0.847 | 22.3 | 0.727 |
| OMP (proposed) | **34.0** | 0.927 | **30.2** | 0.891 | **25.8** | 0.820 |
| CoSamp (proposed) | 33.2 | 0.920 | 29.6 | 0.874 | 24.6 | 0.806 |
| GAP (proposed) | 33.6 | **0.935** | 30.1 | **0.898** | 25.0 | **0.831** |

We find that our proposed algorithms yield better results than state-of-the-art techniques in speckle denoising; especially when the signal is very noisy (poor signal to noise ratio). Although the prior techniques show reasonable PSNR (20+) for all noise levels, the actual perceptual image quality is poor - as can be verified from the SSIM values. To corroborate our claim, we show some denoised images in Fig. 1. Our proposed methods show very similar results; therefore we show the results from one of them only (owing to limitations in space).

The denoised images support the numerical (SSIM) results. The results are shown for the scenario where the input noise level corresponds to a PSNR of 10. The prior techniques yield blurred images, whereas our method yields decent denoising. The blurring owes to the fact that the prior methods assume the logarithm of the image to be sparse; this smoothes out the signal at the onset, hence the denoising results is also smooth. Our method assumes the image to be sparse in the transform domain - an assumption widely tested in denoising and Compressed Sensing based reconstruction. There is no smoothing transform (like logarithm) prior to the application of the sparsifying transform; hence the denoised output is sharper.

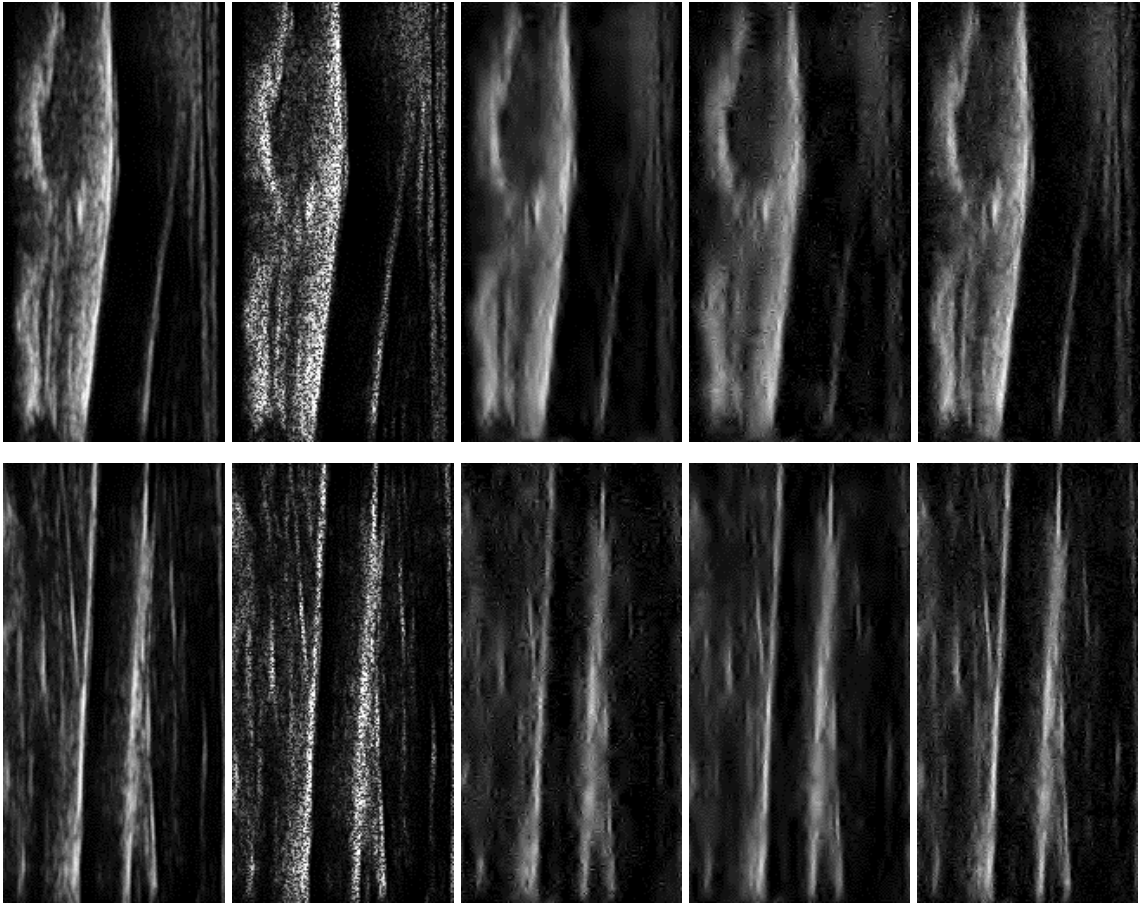

**Figure 4.** Left to Right - Clean Image, Noisy Image, Wavelet + NLM [6], Framelet Diffusion [16], Proposed non-linear OMP.

## 6. Conclusion

In this work, we modify popular greedy sparse recovery algorithms to solve non-linear problems. The OMP and the CoSaMP are modified to solve the synthesis prior problem and the GAP is modified for solving the analysis prior problem. We have run standard tests to evaluate the performance of these algorithms; we find that the trends (variation of success rate with sampling ratio, variation of success rate with sparsity level) are similar to the ones obtained for the linear case.

We apply these algorithms to the problem of speckle denoising in ultrasound images. The results show that the proposed techniques yield better results than state-of-the-art methods compared against. Denoising is a simple problem considering the fact that the problem is fully determined. Our algorithms are capable of solving under-determined problems as well.